\documentclass[aps,prb,superscriptaddress,preprint]{revtex4}
\usepackage{amssymb}
\usepackage{yfonts}
\usepackage{graphicx}
\usepackage{amsmath}
\usepackage{color}
\usepackage{epstopdf}
\usepackage{ulem}

\begin{document}

\title{Thermal fluctuations in antiferromagnetic nanostructures}
\author{Yuriy G. Semenov}
\affiliation{Department of Electrical and Computer Engineering,
North Carolina State University, Raleigh, NC 27695, USA}
\author{Xiniy Xu}
\affiliation{Department of Electrical and Computer Engineering,
North Carolina State University, Raleigh, NC 27695, USA}
\author{Ki Wook Kim}
\affiliation{Department of Electrical and Computer Engineering,
North Carolina State University, Raleigh, NC 27695, USA}
\affiliation{Department of Physics, North Carolina State University,
Raleigh, NC 27695, USA} \email{kwk@ncsu.edu}

\begin{abstract}
A theoretical model is developed that can accurately analyze the effects of
thermal fluctuations in antiferromagnetic (AFM) nano-particles. The approach
is based on Fourier series representation of the random effective field with
cut-off frequencies of physical origin at low and high limits while
satisfying the fluctuation-dissipation theorem at the same time. When
coupled with the formalism of a Langevin dynamical equation, it can describe
the stochastic N\'{e}el vector dynamics with the AFM parameters,
circumventing the arbitrariness of the commonly used treatments in the
micro-magnetic simulations. Subsequent application of the model to
spontaneous N\'{e}el vector switching provides a thermal stability analysis
of the AFM states. The numerical simulation shows that the AFM states are
much less prone to the thermally induced accidental flips than the
ferromagnetic counterparts, suggesting a longer retention time for the
former.
\end{abstract}







\maketitle

\section{Introduction}

Thermal fluctuations are evidently considered a destructive factor as the
electronic devices shrink to the nanoscale dimensions. However, the
situation is not so clear cut in spintronic devices based on magnetic
switching in nano-particles. While spontaneous reversal of the magnetization
negatively affects the lifetime of the binary states in a magnetic memory or
logic, thermal magnetic fluctuations can be exploited to accelerate or even
determine the magnetization flip in ferromagnetic (FM) devices for
heat-assisted magnetic recording.~\cite%
{Berkov2004,Atxitia2009,Galatsis2009,Vogler2016} Furthermore, other widely
adopted mechanisms such as spin transfer torque require a slight
misalignment for the desired magnetization rotation.~\cite{Slavin2009}

The theoretical approaches to the modeling of thermal effects are based on
stochastic itinerancy in the magnetization orientation induced by a fluctuating
effective field $\mathbf{H}_{th}(t)$ that essentially supposes a white noise
satisfying~\cite{Brown1963}%
\begin{equation}
\left\langle H_{th,j}(t)H_{th,j^{\prime }}(t^{\prime })\right\rangle =D_{%
\mathrm{FM}}\delta _{j,j^{\prime }}\delta (t-t^{\prime }).  \label{fm1}
\end{equation}%
The amplitude of un-correlated random component $j=x,y,z$ of the vector
$\mathbf{H}_{th}(t)$ is regulated by the fluctuation-dissipation
theorem
\begin{equation}
D_{\mathrm{FM}}=\frac{2\alpha_\mathrm{G} k_{B}T}{\gamma M_{\mathrm{FM}}V},  \label{fm2}
\end{equation}%
where $\alpha_\mathrm{G} $, $\gamma $, $M_{\mathrm{FM}}$, and $V$ denote the Gilbert
damping parameter, gyromagnetic ratio, magnetization, and volume of the FM
mono-domain, respectively. Incorporating $\mathbf{H}%
_{th}(t)$ into the stochastic Landau-Lifshitz-Gilbert equation (LLG) in the
form of a white noise, while satisfying the requirement of no correlation at
different times as described [Eq.~(1)], gives rise to the mathematical problem of accounting for
a "rapidly varying, highly irregular function".~\cite{Gardiner} Further, this treatment of
$\mathbf{H}_{th}(t)$ leads to an infinite variance.

In the quantitative studies, micro-magnetic simulations have been used widely that treat
the FM particle as an ensemble of magnetic cells with a FM exchange
interaction between them.~\cite{Magnetism2007} Each cell is subjected to a random field
$\mathbf{H}_{th}(t)$ which is not correlated with those of the neighbors.
Likewise, the random fields in the time-domain implementation are assumed invariable once selected
during each time interval $ \Delta t$ (in the range of 20$-$100 ps)  and without interdependence between
the time steps. This approach recovers a finite variance \cite%
{Magnetism2007}%
\begin{equation}
\left\langle H_{th,j}^{2}(t)\right\rangle =\frac{2\alpha_\mathrm{G} k_{B}T}{\gamma M_{%
\mathrm{FM}}V\Delta t}  \label{fm3}
\end{equation}%
that is also in compliance with the fluctuation-dissipation theorem discussed above.
The arrangement adopted in the time domain is in recognition of the finite auto-correlation time $\tau _{c}$
in the magnetization dynamics of the realistic systems in contrast to the no correlation
assumption of Eq.~(1).~\cite{Magnetism2007}  Nonetheless,
the arbitrariness in the time discretization $\Delta t$ of the stochastic fields adds
a significant uncertainty in the final results.~\cite{Roma2014} Similarly, concerns exist on
applicability of the bulk parameters $\alpha_\mathrm{G} $, $M_{\mathrm{FM}}$ and $V$ to
(sub-)nanometer scale cells with arbitrarily chosen sizes and shapes.


The difficulties of the conventional treatment are compounded in the
simulation of more complex antiferromagnetic (AFM) dynamics. The AFMs have recently received much
attention due to their potential advantages in spintronic applications over
the FM counterparts.~\cite{Jungwirth2016} Accordingly, accurate description
of N\'{e}el vector dynamics is crucial in the realistic conditions at
the ambient temperature. In this work, we develop an alternative approach
for the effect of the random thermal
fluctuations in the AFM structures based on a Langevin-type dynamical
equation. The model is then adopted to analyze thermal stability of the AFM
states in the nano-scale dimensions (i.e., lifetimes) highlighting its
potential applications.

\section{Thermal field modeling in AFMs}

Treating the ratcheted field effect on AFM stochastic dynamics as antiparallelly
ordered FM cells in the
manner of micro-magnetic simulations poses challenges beyond those faced by
FM counterparts. For one, the dynamic equations for AFM (or N\'{e}el) vector
$\mathbf{L}$ ($=\mathbf{M}_{1}-\mathbf{M}_{2}$, where $\mathbf{M}_{1}$ and $%
\mathbf{M}_{2}$ are sublattice magnetizations) involve the time derivative $%
\frac{d}{dt}\mathbf{H}_{th}(t)$ that is not compatible with abrupt
changes in $\mathbf{H}_{th}(t)$ often associated with the thermal fields.
Even when this singularity is somehow avoided, 
such a step-function treatment would evidently overestimate the high
frequency components of the noise, resulting in parasitic excitation of
spurious optical magnons in the AFM. Another difficulty stems from the fact
that the correlation time of thermal fields may be comparable to the
temporal scale of AFM dynamics which is much faster than the FM
counterparts.~\cite{Atxitia2009} In fact, the effect of a finite correlation
time was a subject of a detailed investigation even for much slower FM
dynamics in the case of colored thermal noise.~\cite{Tranchida2016} In
general, an increase in the auto-correlation time would enhance the inertial
effects and lead to stronger magnetization damping. Thus, the desired solution
is a representation of the thermal fields with a finite, physically determined
correlation time that can be incorporated into the stochastic N\'{e}el vector
dynamics for numerical evaluation. An alternative theory based on the Fokker-Planck
equation was proposed previously to describe small deviations around the deterministic
trajectory of the N\'{e}el vector.~\cite{Gomonay2013}  In contrast, the approach
pursued here can lead to an exact solution of the stochastic
equation that is applicable to large fluctuations as well.

The thermal noise model under consideration is based on
a spectral representation of $\mathbf{H}_{th}(t)$ in contrast
to the introduction of random step functions in the time domain as it allows
a number of advantages. First, the response of a damped AFM vector indicates
that the random fluctuations via $\mathbf{H}_{th}(t)$  also decay within the
corresponding characteristic time $\tau_{m}$ (i.e., the longest time of relevance).
Thus, $2 \pi/\tau_{m}$ essentially provides the truncation frequency
in the noise spectrum.  Similarly, the auto-correlation time $\tau _{c}$
(or more precisely, its inverse) can be incorporated as the upper bound in the high
frequencies. Further, the association of $\tau_{m}^{-1}$ to the broadening
$\delta_r$ of AFM resonant frequency ($\delta_r = 2 \pi/\tau_{m}$) offers a physical
ground for the discretization of the spectral domain in the comparable intervals
$\Delta \omega$ ($\approx \delta_r$). This frequency uncertainty in each interval
conveniently enables us to approximate the desired spectral function with an
average over $\Delta \omega$ around a typical frequency, followed by a summation over
the allowed frequency domain. In other words, the AFM response
on the actual thermal noise in a dissipative medium is virtually equivalent to a series of
harmonic oscillations (i.e., Fourier expansion) with random amplitudes and frequencies $n\Delta \omega$
($n=1,2,...,N$, where $N = 2 \pi/ \tau_c \Delta \omega \approx \tau_m/\tau_c$).
As such, a similar Fourier series treatment can also be used for $H_{th,j}(t)$.

The underlying rationale of the discrete treatment described above can be seen more clearly by considering the
formal conversion of the response to the white noise. Since the magnetic permeability $\chi (\omega )$
(treated here like a scalar for convenience in the notation) drives the dynamical response $ m(\omega )$, the resulting
stochastic motion  is expressed as ($j =x,y,z$):
\begin{equation}
m_j(t)=\int \chi (\omega )H_{th,j}(\omega )e^{i\omega t}d\omega =\sum_{n}\int_{0}^{\Delta
\omega }\chi (n\Delta \omega +\omega )H_{th,j}(n\Delta \omega +\omega
)e^{i(n\Delta \omega +\omega )t}d\omega . \label{disc}
\end{equation}
If $\Delta \omega $ is small enough to keep $\chi (n\Delta \omega +\omega )$ almost constant
as $ \omega $ varies in the interval $(0,\Delta \omega )$, Eq.~(\ref{disc}) can be reduced to
a discrete sum $\sum_{n}\chi (n\Delta \omega
)e^{in\Delta \omega t}H_{jn}$, where
$H_{jn}=\int_{0}^{\Delta \omega }H_{th,j}(n\Delta \omega +\omega
)e^{i\omega t}d\omega $.  Following the same theoretical underpinning, it can also be shown
explicitly that $H_{jn}$ corresponds to the $n$th Fourier component
of $H_{th,j}(t)$ in the complex space: i.e.,
\begin{equation}
H_{th,j}(t)=\int  H_{th,j}(\omega )e^{i\omega t}d\omega =\sum_{n}e^{in\Delta \omega t}\int_{0}^{\Delta
\omega } H_{th,j}(n\Delta \omega +\omega )e^{i \omega t}d\omega = \sum_{n}H_{jn} e^{in\Delta \omega t} . \label{hdisc}
\end{equation}
As the field $H_{th,j}$ is random, so is $H_{jn}$.  With explicit imposition of the upper
and lower bounds in the noise spectrum discussed earlier, the thermal field can finally be written as a
series of harmonic perturbations with random amplitudes $ a_{jn}$ and $b_{jn}$
in the following form:
\begin{equation}
H_{th,j}(t)=\sum_{n=1}^{N}a_{jn}\sin (n\Delta \omega t)
+ \sum_{n=1}^{N}b_{jn}\cos (n\Delta \omega
t).  \label{fht}
\end{equation}%
Note that this noise expression applies only for
a duration up to $2\pi /\Delta \omega $ in the time domain due to the relaxation
(i.e., $\sim \tau_m$).  A time period longer than this interval requires refreshing
the selection of amplitude for each component. Thus, $H_{th,j}(t)$ (as well as
the associated quantities such as the correlation function) is not periodic in time.
Nonetheless, it is continuous in $t$ ensured by a condition imposed on $ a_{jn}$ and $b_{jn}$,
which stems from the stationarity of the random process (see the discussion below).

The approach based on Eq.~(\ref{fht}) avoids unphysical features attributed
to the white noise treatment of Eq.~(\ref{fm1}) including the virtually
constant spectral density at arbitrary small frequencies and the excitation
of very high frequency perturbations (see Fig.~1). It also circumvents the
singularities associated with the derivatives of the step-function representation
of the random thermal field. \cite{Tsiantos2003}
As discussed above, the frequency interval $\Delta \omega $ can be associated with the
broadening $\delta _{r}$ of AFM resonance frequency (e.g.,
$\Delta \omega \approx \delta _{r} = 2 \pi/ \tau_m$),
for which experimental measurements are generally available in the literature.
Similarly,  the estimation of the upper bound (i.e., $\tau_c ^{-1}$) can be
reliably achieved in term of the microscopic theory. An alternative is
to treat $\tau _{c}$ as a phenomenological parameter following the earlier
studies for the corresponding problem in the FM particles.~\cite{Magnetism2007}
Needless to say, both of them (i.e., $\tau_m$ and $\tau_c$)
are also a function of temperature as with other relevant parameters of the AFM (including
damping constant, resonance frequency, etc.). This dependence can be accounted for by simply
adjusting the numerical values according to the the ambient conditions of interest.


On the other hand, the thermal field must satisfy a restriction on the correlation
function imposed by the stationarity of the fluctuations as well; i.e.,
\begin{equation}
g_{j,j^{\prime }}(t^{\prime },t)=\delta _{j,j^{\prime }}\left\langle
H_{th,j}(t^{\prime })H_{th,j}(t)\right\rangle ,  \label{cor}
\end{equation}%
where $\left\langle ...\right\rangle $ is an average over the ensemble of
identical magnetic particles. The function $g_{j,j^{\prime }}(t^{\prime },t)$
that depends on $t^{\prime }-t$ can be calculated in terms of Eq.~(\ref{fht}%
) so long as the random parameters are statistically independent, i.e., $%
\left\langle b_{jn}a_{j^{\prime }n^{\prime }}\right\rangle =0$, $%
\left\langle b_{jn}b_{j^{\prime }n^{\prime }}\right\rangle =\delta
_{j,j^{\prime }}\delta _{n,n^{\prime }}\left\langle b_{n}^{2}\right\rangle $%
, $\left\langle a_{jn}a_{j^{\prime }n^{\prime }}\right\rangle =\delta
_{j,j^{\prime }}\delta _{n,n^{\prime }}\left\langle a_{n}^{2}\right\rangle $
and the premise $\left\langle b_{jn}\right\rangle =\left\langle
a_{jn}\right\rangle =0$. Stationarity of the random process $H_{th}(t)$ also
imposes equality $\left\langle b_{n}^{2}\right\rangle =\left\langle
a_{n}^{2}\right\rangle $ such that Eq.~(\ref{cor}) reduces to
\begin{equation}
g(t^{\prime },t)=\frac{1}{2}\sum_{n=1}^{N}\left( \left\langle
a_{n}^{2}\right\rangle +\left\langle b_{n}^{2}\right\rangle \right) \cos
[\Delta \omega n(t^{\prime }-t)]=g(t^{\prime }-t).  \label{stat}
\end{equation}%
Here, subscript $j$ is omitted for simplicity. In the limit of white noise
(i.e., $\Delta \omega \rightarrow 0$ and $N\rightarrow \infty $), this
equation obviously reproduces the $\delta $-correlation as supposed in Eq.~(%
\ref{fm1}), provided that $\left\langle b_{n}^{2}\right\rangle =\left\langle
a_{n}^{2}\right\rangle $ is a constant. Thus, the value $\frac{1}{2}\left(
\left\langle a_{n}^{2}\right\rangle +\left\langle b_{n}^{2}\right\rangle
\right) $ determines the spectral density of the correlation function in
Eq.~(\ref{stat}):%
\begin{equation}
(H_{th}^{2})_{\omega }=\frac{1}{2\delta _{r}}\sum_{n=1}^{N}(\left\langle
a_{n}^{2}\right\rangle +\left\langle b_{n}^{2}\right\rangle ).  \label{sd}
\end{equation}%
\qquad

The set of parameters $\left\langle b_{n}^{2}\right\rangle $ relates to
magnetic susceptibility $\chi (\omega )=\chi ^{\prime }(\omega )+i\chi
^{\prime \prime }(\omega )$ via the fluctuation-dissipation theorem. In the
limit of high temperature, this theorem prescribes%
\begin{equation}
(H_{th}^{2})_{\omega }=2\frac{k_{B}T}{\hbar \omega }\frac{\hbar \chi
^{\prime \prime }(\omega )}{\left\vert \chi (\omega )\right\vert ^{2}}.
\label{fd}
\end{equation}%
We apply the AFM permeability at zero external field in the form~\cite%
{Andreev1980}
\begin{equation}
\chi (\omega )=-\frac{2}{D(\omega )}V\gamma M_{L}(\gamma H_{an}+i\alpha
_\mathrm{A}\omega );  \label{hi}
\end{equation}%
\begin{equation}
D(\omega )\simeq \omega ^{2}-2\gamma ^{2}H_{ex}H_{an}-2i\alpha _\mathrm{A}\omega
\gamma (H_{ex}+H_{an}),  \label{det}
\end{equation}%
where $M_{L}$ denotes the saturation magnetization ($=\left\vert \mathbf{L}%
\right\vert $ in equilibrium), $H_{ex}$ and $H_{an}$ ($\ll H_{ex}$) stand
for the interlayer exchange field and the anisotropy field, respectively,
and $\alpha _\mathrm{A}$ is a damping constant which is associated
with each AFM sublatice (also related to the resonance width $\delta
_{r}=\alpha _\mathrm{A}\gamma H_{ex}$). The validity around the zero-field
resonance frequency $\omega _{r}=\sqrt{2\gamma ^{2}H_{ex}H_{an}}$ is assumed
for the permeability expression given above.

A straightforward calculation with a sufficiently small $\alpha _\mathrm{A}$
provides the power of the thermal field as
\begin{equation}
(H_{th}^{2})_{\omega }=\frac{2\eta k_{B}T}{\gamma M_{L}V}.  \label{pd}
\end{equation}%
This expression formally resembles the thermal effect in a FM mono-domain
[see Eq.~(\ref{fm2})] so long as the modified AFM damping parameter $\eta $ (%
$=\alpha _\mathrm{A}\sqrt{H_{ex}/H_{an}}$) corresponds to the FM
Gilbert damping constant $\alpha_\mathrm{G}$. Comparison of
Eqs.~(\ref{pd}) and (\ref{sd}) yields
\begin{equation}
\frac{1}{2}(\left\langle b_{n}^{2}\right\rangle +\left\langle
a_{n}^{2}\right\rangle )=\delta _{r}\frac{2\eta k_{B}T}{\gamma M_{L}V}.
\label{abm}
\end{equation}%
It is convenient to generate the Fourier amplitudes $a_{n}$, $b_{n}$ of the
thermal field in terms of the random numbers $\alpha _{n}$, $\beta _{n}$ of
the Gaussian distribution with variance of $1$; i.e.,
\begin{equation}
\frac{1}{2N}\sum_{n=1}^{N}\left\langle \beta _{n}^{2}+\alpha
_{n}^{2}\right\rangle =1.  \label{abd}
\end{equation}%
Consequently, Eq.~(\ref{abm}) imposes relations $a_{n}=\overline{B}\alpha
_{n}$ and $b_{n}=\overline{B}\beta _{n}$ with the scaling
\begin{equation}
\overline{B}=\frac{\delta _{r}}{\gamma }\left( \frac{2k_{B}T}{NKV}\right)
^{1/2}  \label{ab}
\end{equation}%
yielding the thermal field in dimensionless units as%
\begin{equation}
\frac{\gamma H_{th}(t)}{\omega _{r}}=\frac{\delta _{r}}{\omega _{r}}\left(
\frac{2k_{B}T}{NKV}\right) ^{1/2}\left( \sum_{n=1}^{N}\alpha _{n}\sin
n\delta _{r}t+\sum_{n=1}^{N}\beta _{n}\cos n\delta _{r}t\right) ,  \label{ht}
\end{equation}%
where $K=M_{L}H_{an}$ is an anisotropy constant. This expression clearly
gives the derivatives $dH_{th}(t)/dt$ in the form of smooth functions that
can be directly included in the AFM dynamic equation.

\section{Langevin equation}

The thermal field effect on the N\'{e}el vector dynamics can now be modeled
in terms of the Lagrangian derived from the symmetry consideration.~\cite%
{Andreev1980,Ivanov1995} The alternative approach based on the LLG equations
for the coupled sublattice magnetizations $\mathbf{M}_{1}$ and $\mathbf{M}%
_{2}$ in an external field $\mathbf{H}$ (including the contribution $\mathbf{%
H}_{th}$ of thermal origin) as well as the internal exchange and anisotropy
fields ($\mathbf{H}_{ex}$, $\mathbf{H}_{an}$) generates the same result when
the AFM exchange coupling dominates over the others. The latter condition
supposes the magnitude of the N\'{e}el vector $\left\vert \mathbf{L}%
\right\vert$ to remain unaltered under its rotation such that the unit
vector $\mathbf{n=L/}\left\vert \mathbf{L}\right\vert $ is sufficient to
uniquely determine the AFM state. Since the following analysis is limited to
the AFMs of nano-scale sizes, the spatial variation of $\mathbf{L}$ can be
safely omitted in the Lagrangian, which takes the form
\begin{equation}
\textswab{L}=\frac{M_{L}^{2}}{2\omega _{ex}^{2}}{\dot{\mathbf{n}}}^{2}-\frac{%
M_{L}^{2}}{\omega _{ex}^{2}}[{\dot{\mathbf{n}}}\times \mathbf{n]\cdot }%
\gamma \mathbf{H}+\frac{M_{L}^{2}}{2\omega _{ex}^{2}}[\mathbf{n\times }%
\gamma \mathbf{H}]^{2}-W(\mathbf{n}),  \label{Lg}
\end{equation}
where ${\dot{\mathbf{n}}\equiv }\frac{d}{dt}\mathbf{n}$ and $\omega
_{ex}^{2}=\gamma ^{2}H_{ex}{M_{L}}$. We consider the typical case of a
biaxial AFM with the density of anisotropy energy
\begin{equation}
W(\mathbf{n})=\frac{1}{2}(K_{x}n_{x}^{2}+K_{z}n_{z}^{2}),  \label{an}
\end{equation}%
where the constants $K_{x}$ ($<0$) and $K_{z}$ ($>0$) determine the easy $x$%
- and the hard $z$-axis, respectively. In addition, the magnetic anisotropy
can be engineered via the shape and the strain of the AFM sample.~\cite%
{Gomonay2014} The cubic and higher-order terms are neglected in Eq.~(\ref{an}%
). Accordingly, the anisotropy field $H_{an}$ now corresponds to $\left\vert
K_{x}\right\vert /M_{L}$ [i.e., $\left\vert K_{x}\right\vert \leftrightarrow
K$ in Eq.~(\ref{ht})].

Then, the magnetic relaxation toward the local minimum of $W(\mathbf{n},t)$
can be incorporated into the kinetic equation by way of a dissipation
function
\begin{equation}
\textfrak{R}=\frac{\delta _{r}{M_{L}}^{2}}{\omega _{ex}^{2}}{\dot{\mathbf{n}}%
}^{2},  \label{3}
\end{equation}%
which can be given in terms of the homogeneous line width $\delta _{r}$ of
the AFM resonance mentioned earlier. Note that this
expression accounts for only the relativistic Gilbert-like relaxation.
The effect of the exchange relaxation on $\mathbf{n}$ is expected to be relatively unimportant
as it primarily affects the net magnetization of the AFM (i.e., $\mathbf{M}_1 + \mathbf{M}_2$)
rather than the actual dynamics of the N\'{e}el vector ($=\mathbf{M}_1 - \mathbf{M}_2$) [see
Ref.~\onlinecite{Gomonay2014} for a related discussion].

The corresponding Lagrange equation
augmented with Eq.~(\ref{3}) describes the evolution of the AFM vector in
the form of a Langevin second-order differential equation. Since the
variation $\delta \mathbf{n}$ of unit vector $\mathbf{n}$ comes from its
rotation around a vector $\delta \mathbf{\phi }$ by an infinitesimal angle
$\left\vert \delta \mathbf{\phi }\right\vert $, the resulting expression
takes the form
\begin{equation}
\mathbf{n}\times \left[ {\ddot{\mathbf{n}}}-2({\dot{\mathbf{n}}}\times
\mathbf{h})-(\mathbf{n}\times {\dot{\mathbf{h}}})+\mathbf{h}(\mathbf{n\cdot h%
})+\frac{\partial }{\partial \mathbf{n}}w(\mathbf{n})+2\frac{\delta _{r}}{%
\omega _{r}}{\dot{\mathbf{n}}}\right] =0  \label{LE}
\end{equation}%
in dimensionless time $\omega _{r}t\rightarrow t$. Similarly, a normalized
form is used for the field $\mathbf{H}$ (i.e., $\mathbf{h}=\gamma \mathbf{%
\mathbf{H/}}\omega _{r}$). Hereinafter, $\mathbf{h}$ corresponds to the
normalized thermal field $\mathbf{h}_{th}$ assuming no contribution of other
origins.

The actual independent variables are polar $\varphi $ and azimuthal $\theta $
angles of the unit vector $\mathbf{n}=(\sin \theta \cos \varphi ,\sin \theta
\sin \varphi ,\cos \theta )$. Accordingly, Eq.~(\ref{LE}) establishes the
set of two second order differential equations
\begin{eqnarray}
&&\ddot{\theta}=\frac{1}{2}\left[ \dot{\varphi}^{2}+\kappa _{z}-\kappa
_{x}\cos ^{2}\varphi \right] \sin 2\theta -2\lambda \dot{\theta}  \label{TE}
\\
&&+2\dot{\varphi}\sin \theta (\mathbf{n}\cdot \mathbf{h})+(\dot{{h}_{x}}\sin
\varphi -\dot{{h}_{y}}\cos \varphi )+F_{\theta }(\mathbf{h})  \notag
\end{eqnarray}%
and
\begin{eqnarray}
&&\ddot{\varphi}\sin ^{2}\theta =\frac{\kappa _{x}}{2}\sin 2\varphi \sin
^{2}\theta -\dot{\theta}\dot{\varphi}\sin 2\theta -2\lambda \dot{\varphi}%
\sin ^{2}\theta   \label{PE} \\
&&-2\dot{\theta}\sin \theta (\mathbf{n}\cdot \mathbf{h})+\cos \theta (%
\mathbf{n}\cdot \dot{\mathbf{h}})-\dot{{h}_{z}}+F_{\varphi }(\mathbf{h}),
\notag
\end{eqnarray}%
where $\kappa _{z}=K_{z}/\left\vert K_{x}\right\vert $, $\kappa
_{x}=K_{x}/\left\vert K_{x}\right\vert $, and $\lambda =\delta _{r}/\omega
_{r}$. The quadratic-in-$\mathbf{h}$ terms $F_{\theta }(\mathbf{h})$ and $%
F_{\varphi }(\mathbf{h})$ have often been neglected for relatively small
thermal fluctuations around the deterministic N\'{e}el vector traces.~\cite%
{Gomonay2013} In contrast, these two terms cannot be ignored when the
problem concerns spontaneous N\'{e}el vector switching through the barrier
of anisotropy energy. The detailed expressions necessary in the latter case
are given as
\begin{equation}
F_{\theta }(\mathbf{h})=\frac{1}{2}\sin 2\theta \left( -h_{x}^{2}\cos
^{2}\varphi -h_{y}^{2}\sin ^{2}\varphi +h_{z}^{2}\right)   \label{th2}
\end{equation}%
and%
\begin{equation}
F_{\varphi }(\mathbf{h})=\frac{1}{2}\sin ^{2}\theta \sin 2\varphi
(-h_{x}^{2}+h_{y}^{2}).  \label{ph2}
\end{equation}%
Since $\left\langle h_{x}^{2}\right\rangle =\left\langle
h_{y}^{2}\right\rangle =\left\langle h_{z}^{2}\right\rangle $, the thermal
field does not deviate the equilibrium position away from the stationary
states $\mathbf{n}\Vert \pm \hat{\mathbf{x}}$ on average (which still
permits flipping between them). The cross terms $h_{i}h_{j}$ ($i\neq j$) are
dropped safely considering the uncorrelated nature of the fluctuations $h_{i}
$ and $h_{j}$. Note that the stochastic equations given above [e.g., Eqs.~(%
\ref{TE}) and (\ref{PE})] can be readily applied to describe the N\'{e}el
vector dynamics in the presence of the driving field as well as the thermal
fluctuations. In such a case, the field $\mathbf{h}$ (thus, $\mathbf{H}$)
needs to be expanded to include both contributions.
An explicit Runge-Kutt method can be used for the time integration of the
differential equations. The discretization step size for this numerical method depends
on the correlation time, for which a fraction of $\tau _{c }$ is a convenient choice.

Compared to the evolution of FM nano-particle magnetization, the AFM N\'{e}%
el vector dynamics is much more complex due to several reasons. For
instance, the relatively strong fluctuations may disturb the trajectory in
such a manner that does not nudge the N\'{e}el vector out of the initial
stable state. This phenomenon is related to the chiral dynamics of
sublattice magnetizations. Similarly, the inertial behavior can play a
considerable role unlike in the FM counterparts.~\cite{Semenov2017} With
strong damping (i.e., a lesser impact by inertia), one can expect that the N%
\'{e}el vector would be drawn closer to the saddle point of the anisotropy
potential separating two energetically favorable regions. Under slow
relaxation, on the other hand, the nearly free movement with inertia may
migrate away from the saddle point, ultimately requiring a stronger
excitation to overcome the barrier. At the same time, the rate of field
variation (i.e., the slope $\frac{d}{dt}h$) affects the outcome along with
its amplitude [see, for example, Eqs.~(\ref{TE}) and (\ref{PE})].

Nevertheless, the outcome of the stochastic treatment
is expected to mimic the Boltzmann-type thermal distribution in equilibrium.
As a test, a comparison is made in Fig.~2 between the two for
a range of $z$-directional anisotropy values in a biaxial AFM at room temperature.
The result physically corresponds to reorientation of the N\'{e}el vector along
the $z$ direction ($|n_z| \rightarrow 1$) as the primary easy axis of the material
switches from $x$ to $z$ (i.e., $K_z / K_x = 0 \rightarrow 2$).
While Fig.~2(a) plots $100$ independent solutions of stochastic
equations at each $K_z / K_x$ value (after a long but fixed duration $t$),
the data points in Fig.~2(b)
represent the same number of random selections for $\mathbf{n}$ from the Boltzmann
distribution accounting for only the anisotropy energy;
i.e., $\exp [ -\frac{1}{2}(K_{x}n_{x}^{2}+K_{z}n_{z}^{2})/ k_B T]$.
The similarity between them is rather uncanny despite the drastic difference
in the theoretical approaches.  The observed small disparity in the variance may be
attributed to the neglect of the "kinetic" energy in the simple Boltzmann
expression used in Fig.~2(b) [see the first term on the right-hand side in Eq.~(\ref{Lg})].
The non-zero contribution of this (thermal) kinetic energy term tends to reduce the deviation
away from the mean value (i.e., a tighter distribution).

\section{Retention time evaluation}

As an illustration of the ability to describe beyond the small fluctuations
around the deterministic trajectory (e.g., Fig.~2),
the dynamical model discussed above is adopted to study the problem of
spontaneous N\'{e}el vector switching in AFM nanostructures.
Evidently, the stability of a magnetic
state against the thermal excitation is an issue of major significance in
numerous applications of magnetic devices such as nonvolatile logic and
memory. However, a corresponding analysis of the functional dependence in a
parametrically closed form is difficult to achieve as in the theory of
bistable dynamics that is quite sophisticated even for one-dimensional (1D)
classical particles~\cite{Melnikov91} or FM mono-domains.~\cite{Kalmykov2004}
Thus, the results of the Langevin dynamics may be more conveniently
interpreted from an empirical standpoint of a particle escaping from a local
minimum through thermal fluctuations in an open system.~\cite%
{Smelyanskiy1997} A key feature commonly adopted in this context is the
activation law for escape, or inversely, the retention time $t_{r}\sim \exp
(\Delta _{b}/k_{B}T)$. Parameter $\Delta _{b}$ represents the effective
activation energy that depends on the particular energetic profile, the
spectral density of noise, and the correlation time as it was shown for a 1D
classical system with a double-well potential.~\cite{Dykman1990}

To evaluate the escape rate, the numerical solutions are obtained in a
sequence of $N_{i}$ iterations, each with the time interval $\tau _{m}$.
As discussed above in Sec.~II, random selection of the Fourier amplitudes is
refreshed for each iteration by following the thermal noise model, while
the initial state is set by the solution of the preceding time interval. This
sequence is repeated until the number ($N_{sw}$) of observed switching events
between the $n_{x}\approx \pm 1$ states  reach a sufficiently high
value (e.g., a few hundred) to limit the statistical error. Then the
retention time (i.e., the inverse of the escape rate) can be estimated as
\begin{equation}
t_{r}=\frac{\tau _{m}N_{i}}{N_{sw}}.  \label{tr}
\end{equation}%
The expression for $\tau _{m}$ can also be given as $2\pi /\delta _{r}$ in
terms of AFM parameter. In the actual calculation, the values typical for
mono-domain dielectric AFMs such as NiO are used as summarized below:~\cite%
{Khymyn2017} $K_{x}=-2.2\times 10^{5}$~erg/cm$^{3}$, $K_{z}=4.4\times 10^{5}$%
~erg/cm$^{3}$, $H_{an}=630$~Oe, $H_{ex}=9.3\times 10^{6}$ Oe, $2\pi
M_{L}=700 $ Oe, $\gamma =1.76\times 10^{7}$ Oe$^{-1}$s$^{-1}$, and
$\alpha _\mathrm{A}=6\times 10^{-4}$. The corresponding zero-field AFM resonance
frequency $\omega _{r}$ is $2\pi \times 220$~GHz, while the effective line width
$ \lambda$ ($ =\delta_r/\omega_r$) is treated as a variable in the
initial analysis. Clearly $\tau_m$ (thus, $\delta_r$) varies from sample to sample as
it depends on external factors such as the quality of the materials.
Likewise the magnitude of the auto-correlation time $\tau _{c}$
is treated empirically.  Our analysis indicates that the quantity of interest (i.e., the escape
rate) is not significantly affected by the exact value of $\tau_c$ so long as it is
sufficiently shorter than $\tau_m$. As such, a small constant fraction ($ \tau_c = 0.01 \tau_m$)
is assumed in the current calculation for simplicity.  Note also that the temperature
dependence of the AFM material properties listed above is not considered to limit the
parameter space for a clear physical picture.

Figure~3(a) shows the simulation results (data points) obtained for
different values of the AFM damping parameter $\lambda $ and the temperature
$T$. Equation~(\ref{tr}) in combination with the supposed exponential
dependence of $t_{r}$ suggests that $\ln ({N_{i}}/{N_{sw}})$ may be a linear
function of $1/T$. This appears to be clearly the case as the linear fit
matches well with the calculations over a sizable range, leading to an
approximate expression
\begin{equation}
\frac{N_{i}}{N_{sw}}=Ae^{\Delta _{b}/k_{B}T}.  \label{nn}
\end{equation}%
The fact that the stochastic calculations reproduce
the simple Arrhenius activation law provides an additional validation of the
investigated formalism. The effective barrier energy $\Delta _{b}$ and the prefactor $A$
can be readily determined from the slope and the intercept. The extracted $%
\Delta _{b}$ are provided in Fig.~3(b) as a function of $\lambda $. For the specific $T$
and $\lambda$, the retention time in an AFM nano-particle of volume $V$ can be calculated
by multiplying the corresponding $\tau_m$ ($= 2 \pi / \lambda \omega_r$) to the obtained
$N_{i}/N_{sw}$.  The result is shown in Fig.~4 (dashed line) as a function of $V$
for the case of $\lambda = 0.01$ and $T =300$~K.  The lateral dimension (with a square
cross-section $l\times l$) is varied whereas the vertical thickness is fixed at 5~nm.

It is instructive to compare the results with the corresponding values in FM
nano-particles. The retention time in an uniaxial FM has been a subject of
investigation in numerous works in the literature that can be summed up as~%
\cite{Breth2012}%
\begin{equation}
t_{\mathrm{FM}}\simeq \frac{1}{\alpha_\mathrm{G} \gamma H_{an}}\sqrt{\frac{2\pi k_{B}T}{%
KV}}e^{KV/k_{B}T}.  \label{tfm}
\end{equation}%
Estimated values of $t_{\mathrm{FM}}$ are plotted in Fig.~4 as ell by adopting the same
parameters used for the AFMs above and $\alpha_\mathrm{G} =0.01$.
In addition, the result for the AFM with uniaxial symmetry
(blue solid line) is also calculated by setting $K_z =0$  for a more direct
correspondence with the FM case.  While the general shapes are very similar in both AFM and FM
cases, the slopes (thus, the dependence on the volume) are substantially steeper for the
AFMs. Due to the strong exchange field, the AFM states appear to be more robust than the FMs
against the thermal fluctuations except in the very small sizes (e.g., $%
l\lesssim 10$~nm and $17$~nm for the biaxial and uniaxial cases, respectively), where the desirable
non-volatility cannot be achieved. For instance, a retention time of over (or nearly) 10 years
may be realized with an AFM of $30\times 30\times 5$ nm$^{3}$ while the same structure in the
FM phase is expected to be reliable only for a few minutes. As for the comparison in
the ultra-small dimensions, the relative advantage or disadvantage between the AFM and FM structures
cannot be determined reliably due to the limitation of Eq.~(\ref{tfm}).  The validity of this analytical
expression is in question as the estimated $t_\mathrm{FM}$ becomes comparable to the short magnetization
relaxation time (i.e., small $V$).
Between the uniaxial and biaxial AFMs, the latter (i.e., biaxial) structure appears to be more
favorable (or robust).  It is not surprising that lifting of the hard anisotropy axis results in the acceleration
of the escaping rate.



\section{Summary}

A theoretical model is developed to analyze the effects of
thermal fluctuations in the AFM dynamics.
The formalism avoids a number of complications attributed to the
conventional treatment of mimicking an actual AFM with
antiferromagnetically coupled FM cells.~\cite{Li2017}
For example, the latter approach treats the virtual cells as the
real FM particles with intrinsic Gilbert damping parameter,
anisotropy constants, frequency of FM resonance, etc. The lack of
any practical ways to define these parameters in terms of available
experimental methods renders the conventional approach unrealistic.
In contrast, the formalism developed in the present study takes advantage of the AFM
macroscopic parameters and makes it possible to systematically
account for the key characteristics including the correlation time.
Further, the validity of the
approach is not limited to the weak, perturbative effect around the
equilibrium point for it can accurately describe rare events such as
spontaneous switching between quasistable states. Subsequent
application to the thermal stability analysis shows that the AFM
states are substantially less prone to the temperature induced
accidental flips than
the FM counterparts, highlighting a potential advantage of AFM spintronics.~%
\cite{Gomonay2017,Baltz2018}

\begin{acknowledgments}
This work was supported, in part, by the US Army Research Office
(W911NF-16-1-0472).
\end{acknowledgments}

\clearpage

\begin{center}
\begin{figure}[tbp]
\includegraphics[width=12cm,angle=0]{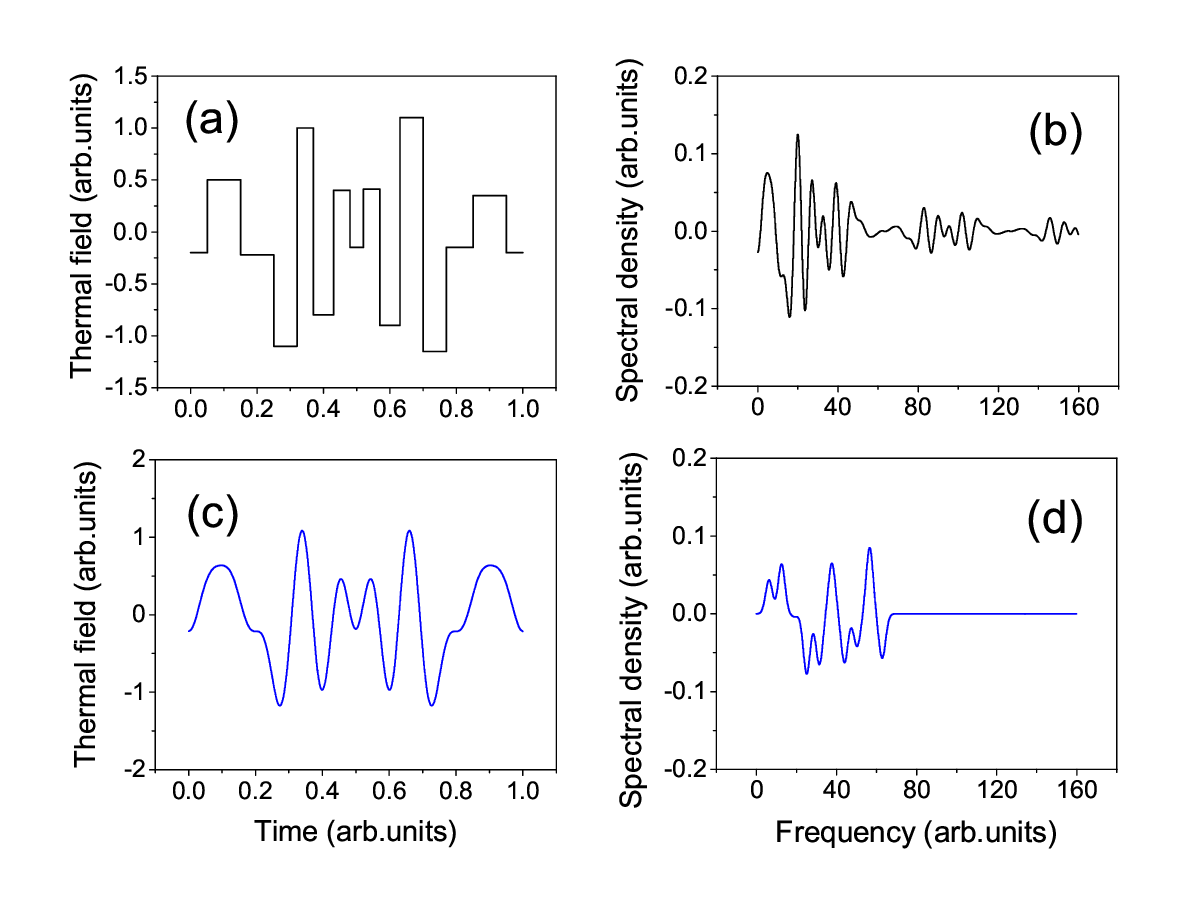}
\caption{Example of white noise simulation in the time domain in terms of
(a) random steps and (c) harmonic oscillations with random amplitudes. While
the corresponding spectral density of the step functions is unrestricted in
the frequency domain [(b)], the harmonics can be confined in a physically
valid range [(d)]. The frequency of each harmonic is assumed to diffuse due
to a finite relaxation time.}
\end{figure}
\end{center}

\clearpage

\begin{center}
\begin{figure}[tbp]
\includegraphics[width=8cm,angle=0]{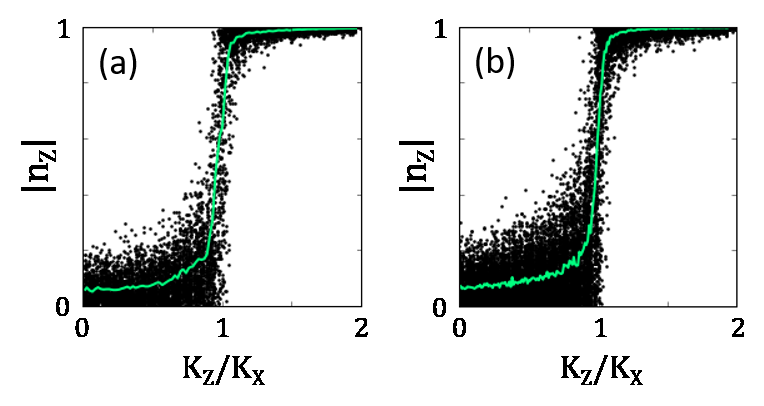}
\caption{Thermal distribution of N\'{e}el vector equilibrium states
(a) calculated in terms of the stochastic equations and
(b) via random selections according to the Boltzmann probability
function. The solid lines indicate the mean values of $n_z$.
Parameters of AFM nano-particle are as discussed in the main text (see Sec.~IV).
While the easy $x$-axis anisotropy ($K_x$) is set at $- 4.4\times 10^{5}$ erg/cm$^{3}$,
the $z$ direction varies from $K_z = 0$ to $K_{z}= -8.8\times 10^{5}$ erg/cm$^{3}$
making it the primary easy axis.}
\end{figure}
\end{center}

\clearpage

\begin{center}
\begin{figure}[tbp]
\includegraphics[width=12cm,angle=0]{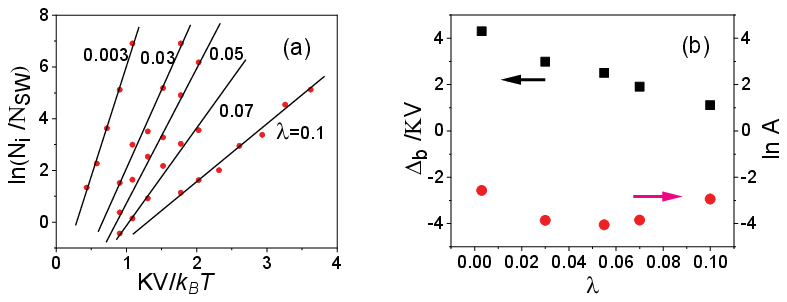}
\caption{
(a) Stochastic simulations of spontaneous N\'{e}el vector reversal [%
$\ln (N_i/N_{sw})$] as a function of the inverse temperature for different
values of the damping constants $\protect\lambda$. The date points (dots)
show the results of the calculation while their exponential approximations
are given by the fitted lines. For sufficient statistics, the iterative
process based on random Monte Carlo selection continues until the number of
observed switching events $N_{sw}$ reaches 150 or more. (b) Effective
barrier $\Delta_b$ and prefactor $A$ determined from the slopes and the
intercepts of the fitted lines shown in (a). Note that $k_B T$ and $\Delta_b$
are normalized to the anisotropy energy $K V$ at the saddle point (where $%
K=\left\vert K_{x}\right\vert $ and $V$ denotes the AFM nano-particle
volume).}
\end{figure}
\end{center}

\clearpage

\begin{center}
\begin{figure}[tbp]
\includegraphics[width=8cm,angle=0]{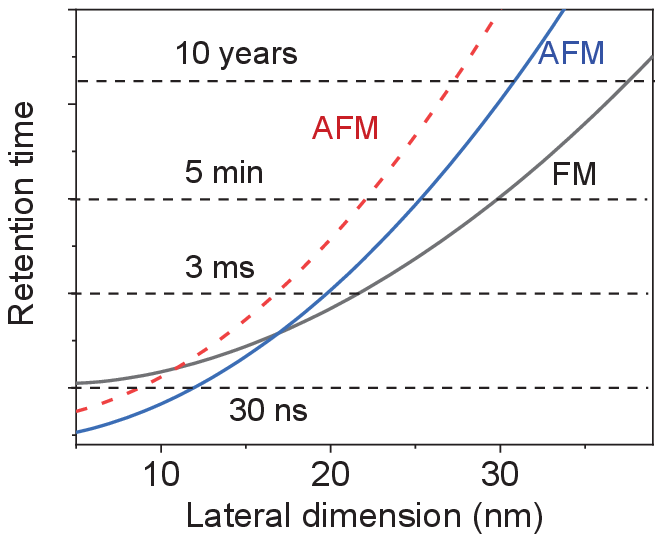}
\caption{
Retention times of AFM and FM nano-particles at room temperature as a function
of the lateral dimension. The solid lines represent the cases of uniaxial symmetry
(both AFM and FM), while the dashed line is for the biaxial AFM nano-particle
($K_x < 0$, $K_z > 0$). A square cross-section is assumed whereas the
vertical thickness is fixed at 5~nm. }
\end{figure}
\end{center}

\end{document}